\documentclass[prb,showpacs]{revtex4}
\usepackage{graphicx,psfrag,amssymb,amsmath}
\begin{document}
\bibliographystyle{apsrev}
\title{Gapped optical excitations from gapless phases: imperfect nesting in unconventional density waves}
\author{Bal\'azs D\'ora}
\email{dora@kapica.phy.bme.hu}
\author{Andr\'as V\'anyolos}
\affiliation{Department of Physics, Budapest University of Technology and 
Economics, H-1521 Budapest, Hungary}
\author{Kazumi Maki}
\affiliation{Department of Physics and Astronomy, University of Southern
California, Los Angeles CA 90089-0484, USA}
\author{Attila Virosztek}
\affiliation{Department of Physics, Budapest University of Technology and 
Economics, H-1521 Budapest, Hungary}
\affiliation{Research Institute for Solid State Physics and Optics, P.O.Box
49, H-1525 Budapest, Hungary}

\date{\today}

\begin{abstract}
We consider the effect of imperfect nesting in quasi-one-dimensional unconventional density
waves in the case, when the imperfect nesting and the gap depends on the same wavevector component.
 The phase diagram is very similar to that
in a conventional density wave. 
The density of states is highly asymmetric with respect to the Fermi
energy.
 The optical conductivity at 
$T=0$ remains unchanged for small deviations from perfect nesting. For higher imperfect nesting parameter, 
an optical gap opens, and considerable amount of spectral weight is
transferred to higher frequencies. This makes the optical response of our
system very similar to that of a conventional density wave.
Qualitatively similar results are expected in d-density waves.
\end{abstract}


\maketitle

\section{Introduction}

The basic ingredient of the density wave (DW) formation is a band structure
consisting of a pair of Fermi sheets, which can be nested to each other
with a certain wavevector,
giving rise to the density wave instability\cite{gruner}. In real materials, however,
this condition is not perfectly fulfilled: $\varepsilon({\bf
k})+\varepsilon({\bf k-Q})\neq 0$. In quasi-one-dimensional models studied during the early history of DW, 
one
 can choose it
as $\varepsilon({\bf k})+\varepsilon({\bf
k-Q})=2\epsilon_0\cos(2bk_y)$, which shows the deviation from the one
dimensionality\cite{ishiguro,mihaly}. In conventional CDW such as NbSe$_3$, the depression of the 
transition
temperature under pressure is described in terms of the pressure dependence
of imperfect nesting, and the large ratio of $2\Delta/T_c$ is also
interpreted\cite{yamaji1,yamaji2,huang0}. Similarly in field-induced SDW many features are successfully
described by this model\cite{ishiguro}. The general consequence of $\epsilon_0$ is the 
destruction of the density wave phase: imperfect nesting depresses
the DW transition temperature and destroys completely the density wave when
$\epsilon_0$ becomes larger than a critical value. 
Also the imperfect nesting term gives rise to dip structures
in the angular dependent magnetoresistance in $\alpha$-(BEDT-TTF)$_2$KHg(SCN)$_4$\cite{admrprl}
and Bechgaard salts (TMTSF)$_2$PF$_6$\cite{tmtsf}.
This
motivates us to incorporate the effect of imperfect nesting in
unconventional density wave (UDW) theory. UDW is a density wave, whose gap function
depends on the wavevector, vanishes on certain points of the Fermi surface,
allowing for low energy excitations. The average of the gap over the Fermi
surface is zero, causing the lack of periodic modulation of the charge
and spin density. Such systems have been studied and proposed over the years in a
variety of systems\cite{castroneto,Ners1}. These include heavy fermions like 
URu$_2$Si$_2$\cite{IO,roma,GG}, 
CeCoIn$_5$\cite{cecoin}, organic conductors as $\alpha$-(BEDT-TTF)$_2$KHg(SCN)$_4$\cite{mplb} and 
(TMTSF)$_2$PF$_6$\cite{tmtsf}, high
$T_c$ superconductors\cite{nayak,benfatto,capnernst}. 
Two different models are possible: 2D or 3D when the gap and the imperfect
nesting depends on the same or different wavevector component,
respectively. Previously we have analyzed the properties of the 3D model\cite{imperfect}, and now we turn 
to the investigation of the 2D one.

The object of the present paper is to extend the analysis of Refs. \onlinecite{nagycikk,imperfect}
to the presence of imperfect nesting when the gap and the imperfect
nesting depends on the same wavevector component. We discuss the temperature
dependence of the order parameter for different $\epsilon_0$'s.
 The phase boundary is almost the same as in a conventional
DW. The chemical potential is shifted from its original value of
the metallic state due to the presence of imperfect nesting.
The temperature dependence of the order parameter, $\Delta(T,\epsilon_0)$
is anomalous: although it decreases monotonically with increasing
temperature, but exhibits a sharp cusp at $\Delta(T,\epsilon_0)=2\epsilon_0$.
In
the density of states (DOS) the particle-hole symmetry is broken for the 2D model, leading to
asymmetric density of states with respect to the Fermi energy. For high values of $\epsilon_0$, the 
zero of the density of states at the Fermi energy disappears, and DOS becomes finite for all energies.
The optical conductivity is not affected by the deterioration of perfect nesting in a wide parameter range.
By further increasing $\epsilon_0$, the divergent peak at $2\Delta$ is divided into
two new peaks. Moreover, a finite optical gap shows up at $T=0$ in spite of the finite density of states.
Similar behaviour was identified in  two dimensional UDW (the so-called d-density wave\cite{nayak}):  
deviations from perfect nesting induce a finite optical gap\cite{zeyher}.
In clean systems, the weight of the 
Dirac delta peak at zero frequency is finite for all temperatures. 
We expect similar results in d-density waves as well.

\section{Phase diagram}

\begin{figure}[h!]
\psfrag{y}[b][t][1][0]{$T/T_{c0}$, $\Delta(0,\epsilon_0)/\Delta_{00}$}
\psfrag{x}[t][b][1][0]{$\epsilon_0/\Delta_{00}$}
\centering{\includegraphics[width=11cm,height=7cm]{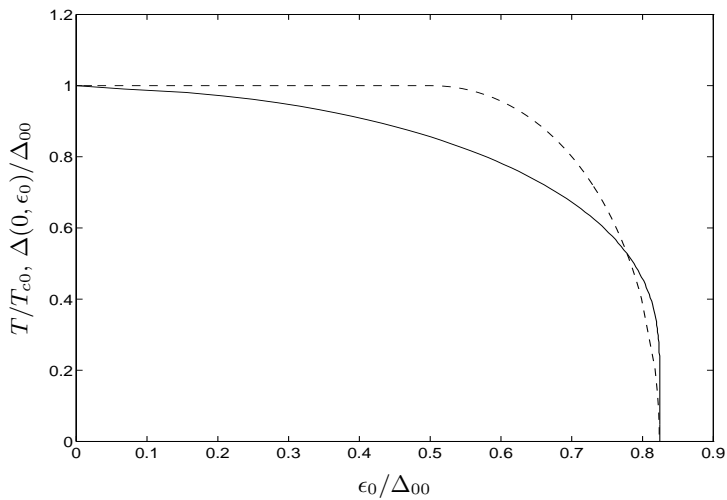}}
\caption{The phase diagram (solid line) and the zero temperature order
parameter (dashed line) are  plotted in the presence of imperfect nesting.\label{fig:fazis2d} }
\end{figure}

\begin{figure}[h!]
%\psfrag{x}[t][b][1][0]{$a(k_x-k_F)$}
\psfrag{y}[t][b][1][0]{$bk_y$}
\psfrag{z}[b][t][1][0]{$E({\bf k})/\Delta$}
{\includegraphics[width=8cm,height=8cm]{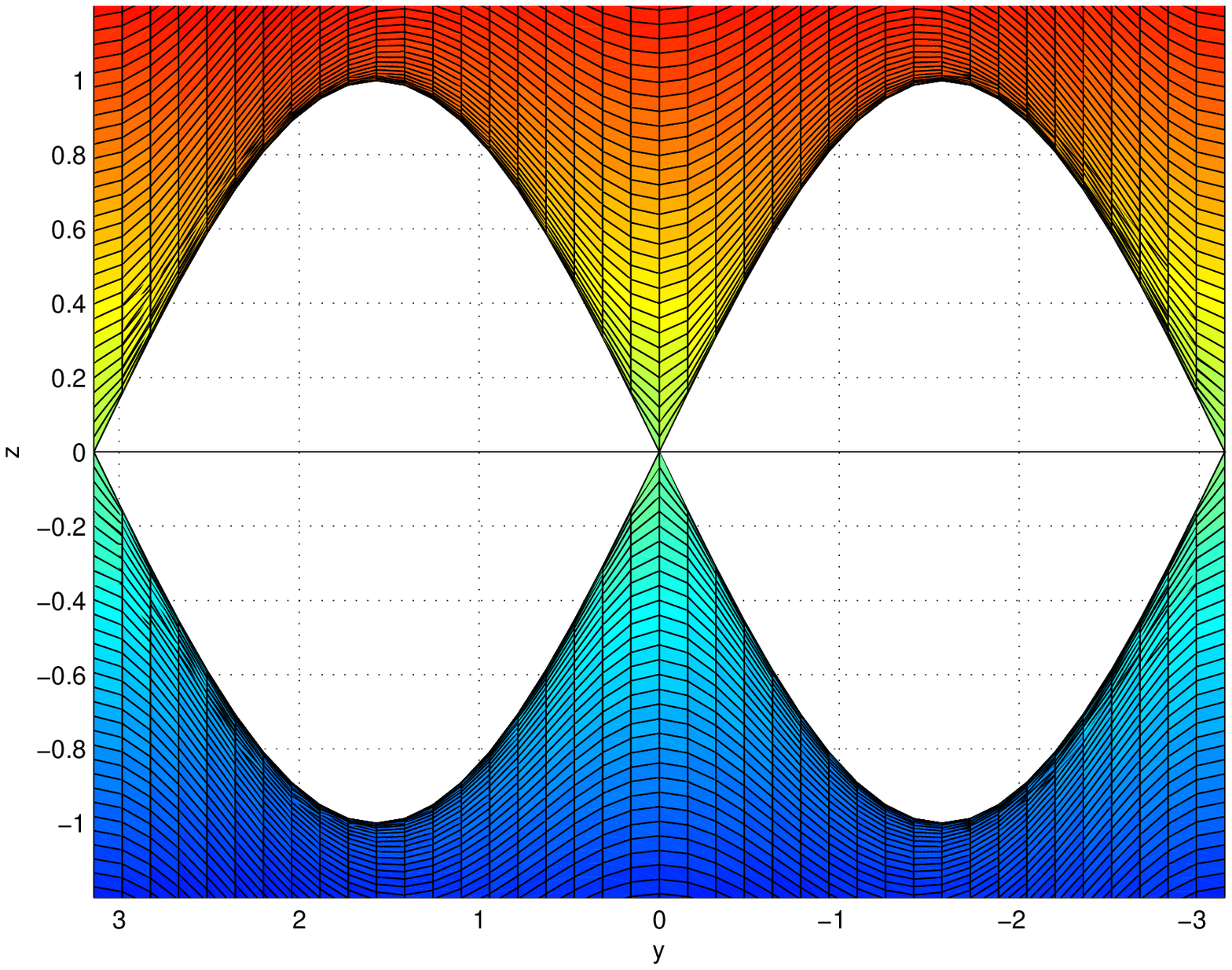}}
\hspace*{4mm}
{\includegraphics[width=8cm,height=8cm]{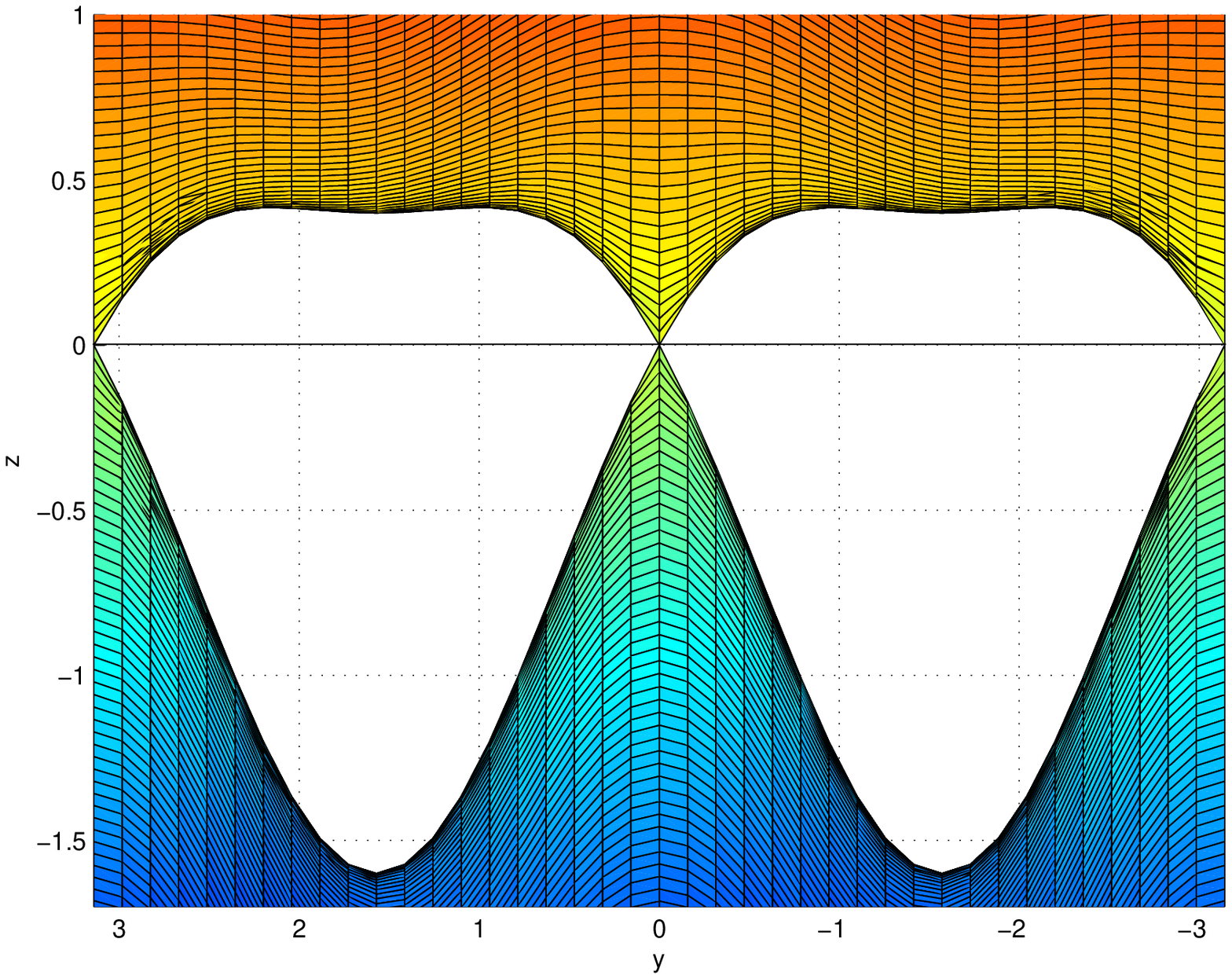}}

\vspace*{9mm}
{\includegraphics[width=8cm,height=8cm]{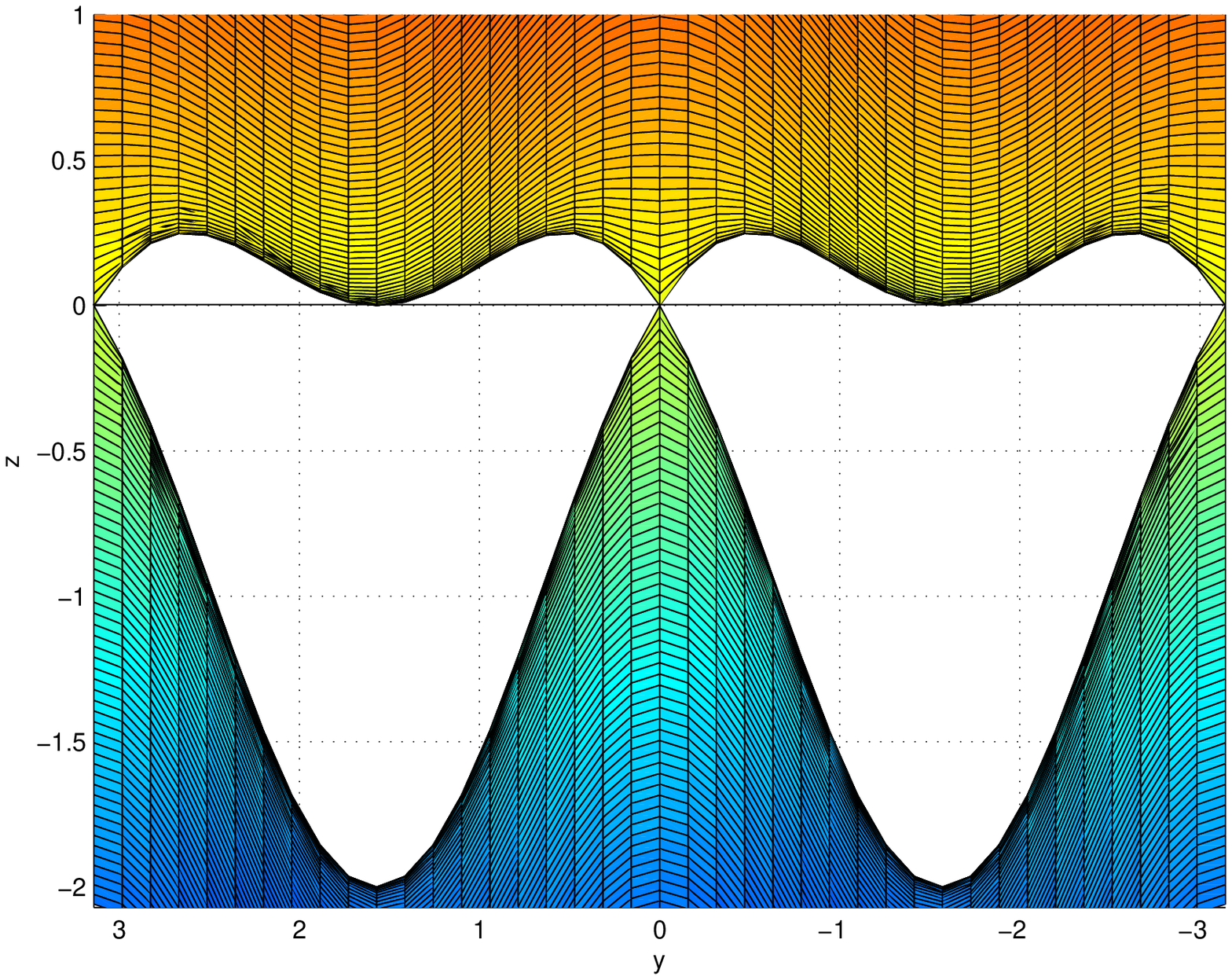}}
\hspace*{4mm}
{\includegraphics[width=8cm,height=8cm]{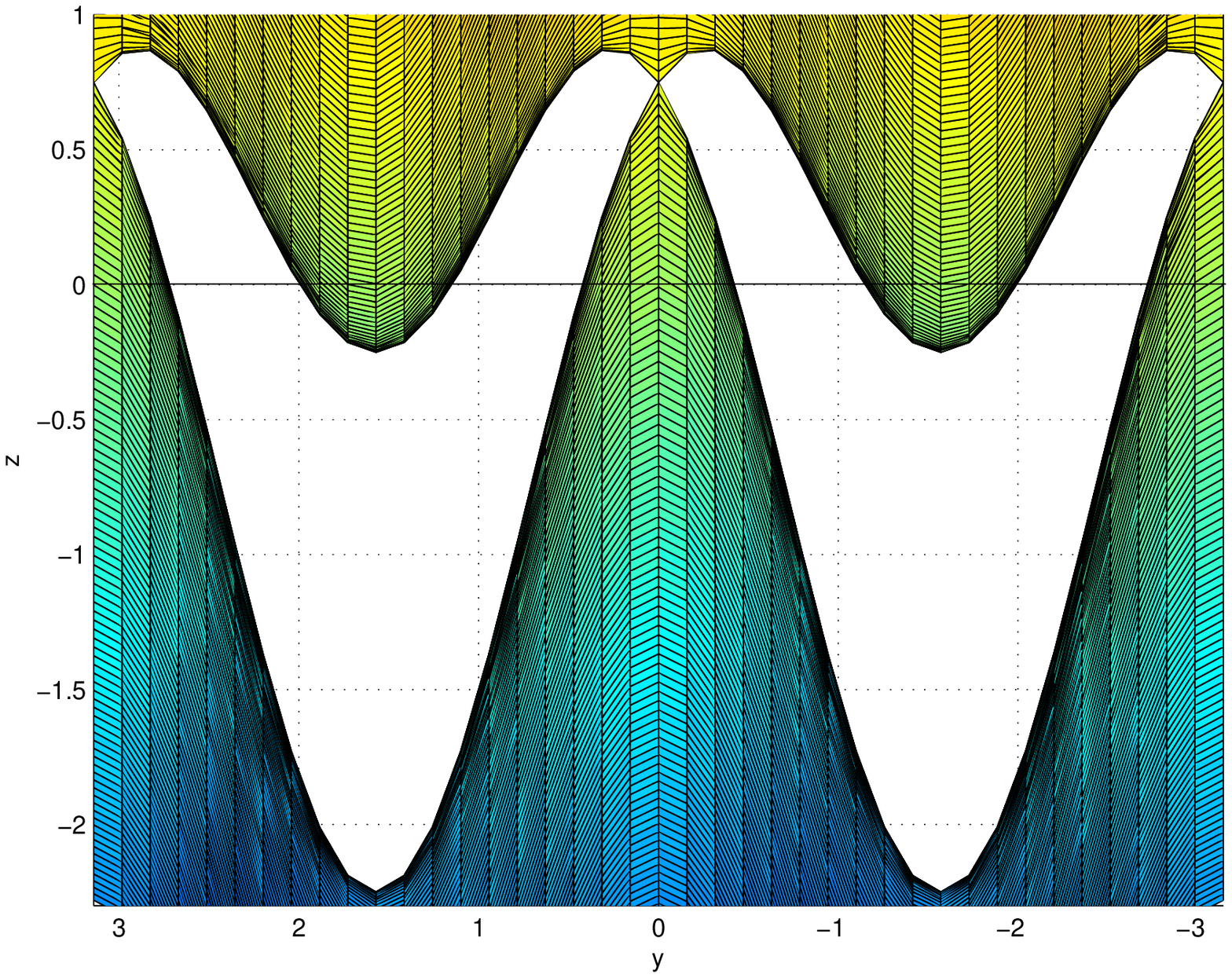}}

\caption{The evolution of the quasiparticle spectrum is shown, viewed from the direction of the $k_x$ axis,  
for $\Delta({\bf k})=\Delta\sin(bk_y)$ in the 
presence of imperfect 
nesting for $\epsilon_0/\Delta=0$, $0.3$, $0.5$ and $1$ from left to right, top to bottom. The horizontal
line denotes the Fermi energy.
The band structure is chosen as $\varepsilon({\bf k})=-2t_a\cos(ak_x)-2t_b
\cos(bk_y)+\varepsilon_0\cos(2bk_y)$ with parameters as $t_a/\Delta=2$, $t_b/\Delta=0.1$ at half filling.\label{fig:spec} }
\end{figure}

The single-particle electron thermal Green's function of UDW is given by\cite{impur,greendw}
\begin{equation}
G^{-1}({\bf k}, i\omega_n)=i\omega_n-\xi({\bf k})\rho_3-\rho_1\sigma_3\textmd{Re}\Delta({\bf
k})-\rho_2\sigma_3\textmd{Im}\Delta({\bf k}),\label{Green0}
\end{equation}
where $\rho_i$ and $\sigma_i$ ($i=1,2,3$) are the Pauli matrices acting on momentum and spin space, 
respectively, and for UCDW
$\sigma_3$ should be replaced by $1$. $\Delta({\bf k})=\Delta e^{i\phi}\cos(bk_y)$ 
or $\sin(bk_y)$. $\phi$ is the unrestricted phase (due to 
incommensurability) of the density
wave, but its explicit value turns out to be irrelevant for our discussion, $\xi(\bf k)$ is the kinetic 
energy spectrum, $\omega_n$ is the fermionic Matsubara frequency.
The effect of imperfect nesting is incorporated in the theory by replacing
the Matsubara frequency in the single particle Green's function with
$\omega_n+i(\epsilon_0\cos(2bk_y)-\delta\mu)$\cite{huang1,huang2}, where
$\delta\mu$ is the change of the chemical potential due to the change in the
spectrum. The order parameter\cite{impur} is assumed to depend on the wavevector like $\Delta({\bf 
k})=\Delta\sin(bk_y)$ or $\Delta\cos(bk_y)$.
The second order phase boundary is given by $\epsilon_0=\Delta_0(T_c)$, 
where $\Delta_0(T)$ is the temperature dependence of the gap in a perfectly 
nested conventional DW with $T_{c0}$ transition temperature. 
$T_{c0}$ is the transition temperature in the absence of imperfect nesting. 
This is almost the complete phase diagram. At high
temperature when $T$ becomes of order of $\epsilon_0$ the deviation from
perfect nesting becomes irrelevant and the best nesting vector is ${\bf
Q}=(2k_F,\pi/b,\pi/c)$. In the conventional scenario two DW phases can occur\cite{hasegawa}, characterized by
slightly different wave vectors and $\bf Q$ is replaced by a temperature
dependent wave vector, opening a narrow region 
above the critical nesting at low temperatures. 
 For the present model, the possibility of ordering with  different
wave vector is there, although its examination is beyond the scope of the
present discussion. The critical nesting
is given by $\epsilon_0=\sqrt e\Delta_{00}/2\approx0.82\Delta_{00}$, where
$\Delta_{00}$ is the gap in a perfectly nested system at zero
temperature. The order parameter remains unchanged for
$\epsilon_0<\Delta_{00}/2$, and vanishes sharply as $\epsilon_0$
approaches its critical value. 
This together with the phase diagram is shown in Fig. \ref{fig:fazis2d}. 
The most interesting consequence of imperfect nesting is that the chemical potential does not 
remain constant under the density wave formation. Its shift is given by
$\delta\mu=\epsilon_0\Theta(\Delta(T,\epsilon_0)-2\epsilon_0)+\Delta
(T,\epsilon_0)^2/(4\epsilon_0)\Theta(2\epsilon_0-\Delta(T,\epsilon_0))$,
where $\Theta(x)$ is the heaviside function. This behaviour can readily
be seen from the density of states, where for any finite $\Delta$ the  
total number of states below the
Fermi energy is regained only by shifting the Fermi energy as
given above. Note that this change belongs to a sinusoidal gap
while for a cosinusoidal gap the sign of the shift is reversed. 
The change in the spectrum in the presence of imperfect nesting is shown in Fig. \ref{fig:spec}, which is given by
\begin{equation}
E_\pm({\bf k})=\frac{\varepsilon({\bf k})+\varepsilon({\bf k-Q})}{2}\pm\sqrt{\left(
\frac{\varepsilon({\bf k})-\varepsilon({\bf k-Q})}{2}\right)^2+|\Delta({\bf k})|^2}.
\end{equation}
In the
perfectly nested case, the low energy part of the spectrum consists of Dirac cones with peaks at the Fermi 
energy\cite{nagycikk}. 
For small $\epsilon_0$, the spectrum is still crossed by the Fermi energy at the zeros of the gap. By 
increasing 
$\epsilon_0$, a broad bump develops in the upper band, and crosses the Fermi energy. At this point, a large 
number of possible states becomes available, and the chemical potential  starts decreasing to 
keep the total number of particles unchanged.

\begin{figure}[h!]
\psfrag{x}[t][b][1][0]{$T/T_{c0}$}
\psfrag{y}[b][t][1][0]{$\Delta(T,\epsilon_0)/\Delta_{00}$}
\centering{\includegraphics[width=11cm,height=7cm]{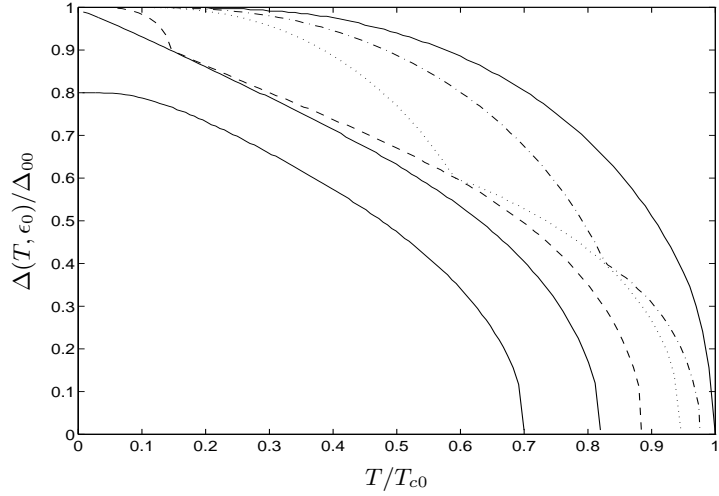}}
\caption{The temperature dependence of
the order parameter for the 2D model is shown for $\epsilon_0/\Delta_{00}=0$,
$0.2$, $0.3$, $0.45$, $0.55$ and $0.7$ from right to left. The cusp shows
up only for $2\epsilon_0<\Delta_{00}$.\label{fig:delta2d} }
\end{figure}

A direct
consequence of this shift is a cusp in the temperature dependence of
$\Delta$ at $\Delta=2\epsilon_0$, since at this point the chemical
potential changes. This feature is shown in Fig. \ref{fig:delta2d}, which is obtained
from the numerical solution of the gap equation:
\begin{equation}
1=T V
\frac{N_0}{4}\sum_n\int\limits_0^{2\pi}\frac{\sin^2(y)dy}{\sqrt{(\omega_n+i(\epsilon_0\cos(2y)-\delta\mu))^2
+\Delta^2\sin^2(y)}},
\end{equation}
where $V>0$ is the interaction responsible for the UDW formation, $N_0$ is
the density of states in the normal state at the Fermi energy per spin.

\section{Density of states}
 
The quasi-particle density of states is given by
\begin{equation}
g_{2D}(E)=N(x,a)=N_0\int\limits_0^{2\pi}\frac{dy}{2\pi}\textmd{Re}
\frac{|E+\delta\mu-\epsilon_0\cos(2y)|}{\sqrt{(E+\delta\mu-\epsilon_0\cos(2y))^2-
\Delta^2\sin^2(y)}}.
\end{equation}
The energy variables are
expressed in units of $\epsilon_0$, i.e. 
$a=\Delta/\epsilon_0$ and $x=(E+\delta\mu)/\epsilon_0$, the energy is
measured from the new Fermi energy. The
density of states is obtained as 
\begin{gather}
N(x,a)=N_0\frac{1}{\pi\sqrt{pq}}\left(\left(x-1-\frac{2q}{p-q}\right)K\left(\frac12
\sqrt{\frac{1-(p-q)^2}{pq}}\right)+\frac{p+q}{p-q}\Pi\left(\frac{(p-q)^2}{-4pq},\frac12
\sqrt{\frac{1-(p-q)^2}{pq}}\right)\right)
\end{gather}
for $x>a^2/8+1$, where $p=\sqrt{(m-1)^2+n^2}$, $q=\sqrt{m^2+n^2}$, $m=(a^2-4(x-1))/8$, $n=a\sqrt{-a^2+8(x-1)}/8$,
$K(z)$ and $\Pi(n,z)$ are the complete elliptic integrals of the first and third
kind\cite{elliptic}. In the remaining regions the DOS reads as
%\begin{subequations}
\begin{eqnarray}
N(x,a)=N_0(\Theta(a-4)f_1(x,a)+\Theta(4-a)f_3(x,a)), &
\dfrac{a^2}{8}+1>x>a-1, \nonumber \\
N(x,a)=N_0\textmd{sgn}(x-1)f_2(x,a), & a-1>x>-a-1,\nonumber \\
N(x,a)=-N_0 f_3(x,a), & -a-1>x, 
\end{eqnarray}
%\end{subequations}
where the following notations are used:
\begin{gather}
f_1(x,a)=\frac{1}{\pi\sqrt{(y_2-1)y_1}}\left((x-1+2y_2)K\left(\sqrt{\frac{y_2-y_1}{(y_2-1)y_1}}\right)
-2y_2\Pi\left(\frac{1}{1-y_2},\sqrt{\frac{y_2-y_1}{(y_2-1)y_1}}\right)\right),
\end{gather}

\begin{gather}
f_2(x,a)=\frac{1}{\pi\sqrt{y_2-y_1}}\left((x-1+2y_2)K\left(\sqrt{\frac{(y_2-1)y_1}{y_2-y_1}}\right)
- 2y_2\Pi\left(\frac{y_1}{y_1-y_2},\sqrt{\frac{(y_2-1)y_1}{y_2-y_1}}\right)\right),
\end{gather}

\begin{gather}                            				
f_3(x,a)=\frac{1}{\pi\sqrt{(1-y_1)y_2}}\left(2(y_2-y_1)\Pi\left(\frac{y_2-1}{y_1-1},
\sqrt{\frac{(1-y_2)y_1}{(1-y_1)y_2}}\right)-2\textmd{sgn}(x-1)\Pi\left(\frac{y_1}{y_1-1},\sqrt{\frac{(1-y_2)y_1}
{(1-y_1)y_2}}\right)\right.+\nonumber\\
+\left.(x-1+2y_1+
\textmd{sgn}(x-1)(x+1))K\left(\sqrt{\frac{(1-y_2)y_1}{(1-y_1)y_2}}\right)
\right),
\end{gather}
and $y_1=(a^2-4(x-1)-a\sqrt{a^2-8(x-1)})/8$,
$y_2=(a^2-4(x-1)+a\sqrt{a^2-8(x-1)})/8$.\\

\begin{figure}[h!]
\psfrag{x}[t][b][1][0]{$E/\Delta$}
\psfrag{y}[b][t][1][0]{$g_{2D}(E)/N_0$}
\includegraphics[width=90mm,height=7cm]{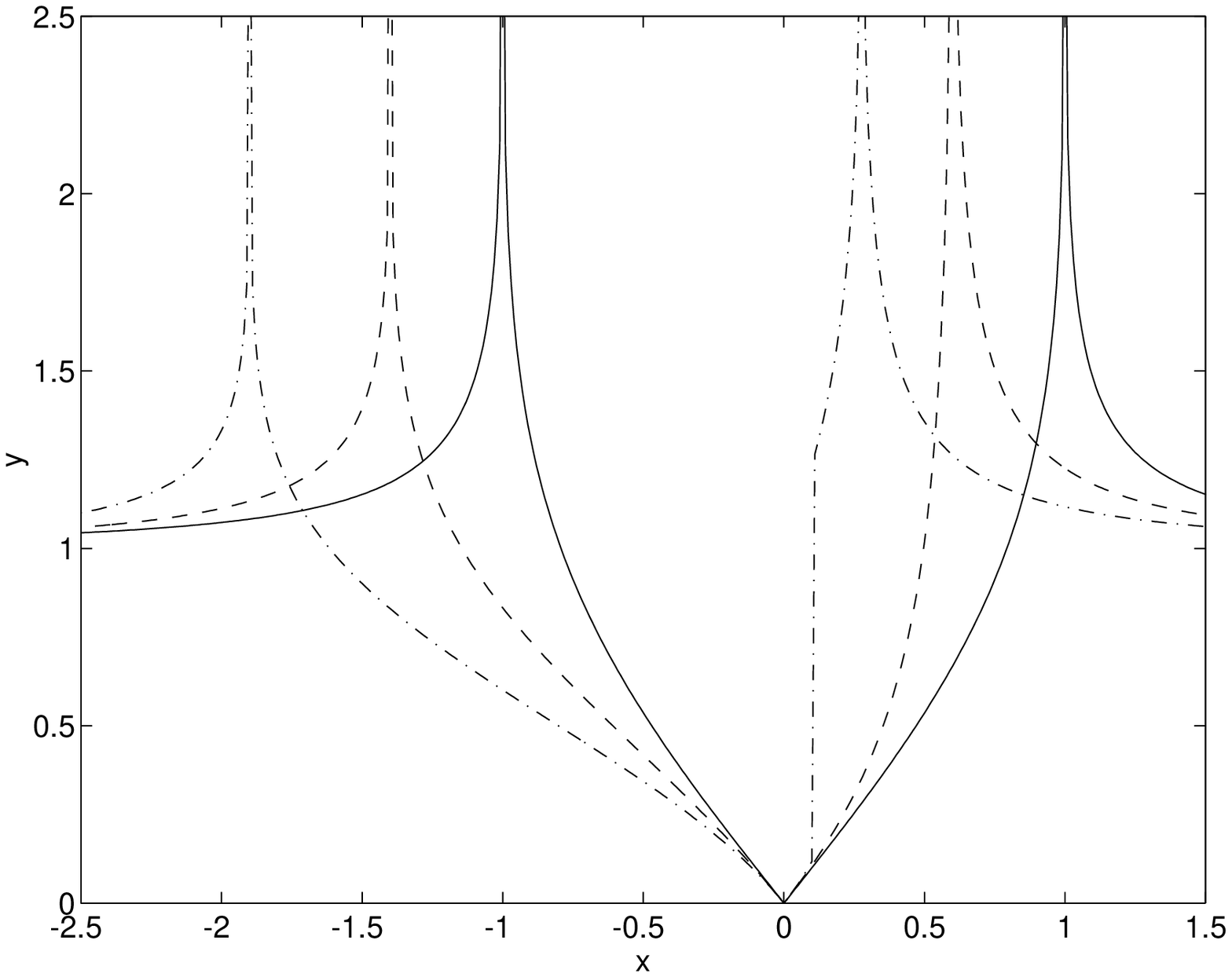}
\hspace*{-10mm}
\includegraphics[width=90mm,height=7cm]{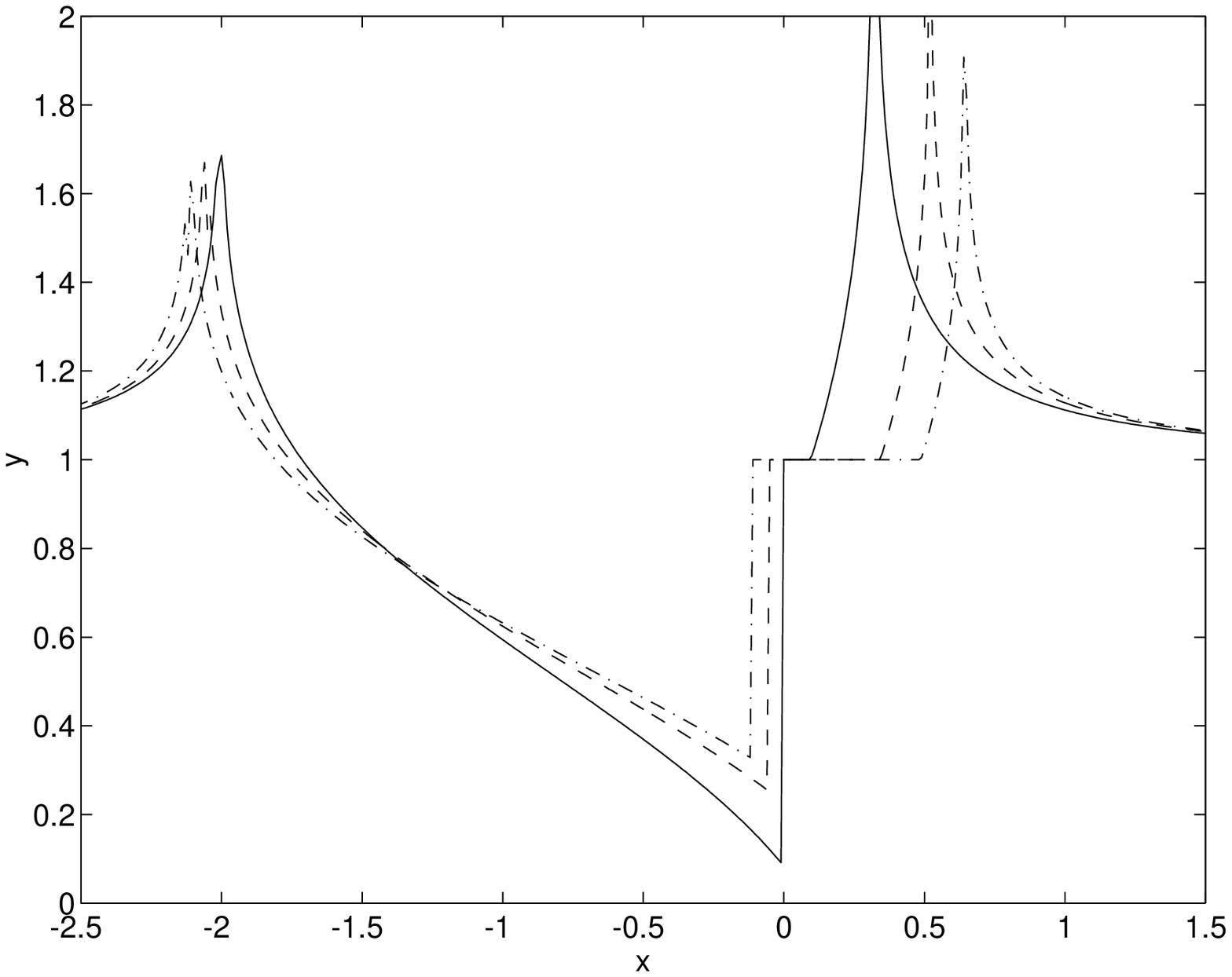}
\caption{The density of states as a function of energy is shown
 in the left panel for $\epsilon_0/\Delta=0$ (solid line), $0.2$ (dashed line), 
$0.45$ (dashed-dotted line). In the right panel $\epsilon_0/\Delta=0.55$ 
(solid line), $0.7$ (dashed line) and $0.8$ (dashed-dotted line) are used.
\label{fig:dos2}}
\end{figure}

The particle-hole symmetry is broken, which can be readily seen from the behaviour of the peaks
in the density of states, which slide from $\pm\Delta$ to
 $-\Delta-\epsilon_0-\delta\mu$ below the Fermi surface, while
above it to $\Delta-\epsilon_0-\delta\mu$ for $4\epsilon_0<\Delta$ and to
$\epsilon_0+\Delta^2/8\epsilon_0-\delta\mu$ otherwise. Also the zero in DOS
is at the new Fermi energy for $\epsilon_0<\Delta/2$, and for larger 
$\epsilon_0$ there exists
no zero in the DOS. The density of states is
plotted in Fig. \ref{fig:dos2}. These statements correspond to
$\Delta({\bf k})=\Delta\sin(k_yb)$,
while for a cosinusoidal gap
$E\rightarrow -E$ change is needed in the density of states. 

The residual density of states
(i.e. $g_{2D}(E=0)$) is given by
$N_0\Theta(2\epsilon_0-\Delta)$. Since on the Fermi surface the DOS
vanishes in the same way for $\epsilon_0\ll\Delta_{00}$ than in the perfectly nested case, 
the specific heat increases quadratically
with temperature close to $T=0$~K in this region, while for large
$\epsilon_0$ it equals to the specific heat in the normal state. 

\section{Optical conductivity}

In this section we investigate the quasiparticle contribution to the
optical conductivity. For simplicity we neglect the effect of the quasiparticle 
damping due to impurity scattering
for example. The quasiparticle part
of the conductivity contains relevant information about the system
in the perpendicular cases ($y$ and $z$) when the effect of the collective contributions
can be neglected. The regular part of the optical conductivity (without the
Dirac delta) is given by
\begin{eqnarray}
\textmd{Re}\sigma_{\alpha\beta}^{reg}(\omega)=N_0\frac{\pi e^2}{\omega^2}\int_{-\pi}^{\pi}\frac{d(bk_y)}{2\pi}\int_{-\pi}^{\pi}\frac{d(ck_z)}{2\pi}\textmd{Re}\frac{v_\alpha({\bf
k})v_\beta({\bf k})\Delta^2({\bf k})}{\sqrt{(\omega/2)^2-\Delta^2({\bf
k})}}\left(\tanh\left(\frac{|\omega|-2\eta}{4T}\right)+\tanh\left(\frac{|\omega|+2\eta}{4T}\right)\right),
\end{eqnarray}
where $v_\alpha({\bf k})$ is the quasiparticle velocity in the $\alpha$
direction, , $v_x=v_F$, $v_y=\sqrt 2t_bb$, $v_z=\sqrt 2 t_cc$ and $\eta=\epsilon_0\cos(2bk_y)-\delta\mu$. From now on we restrict our
investigation to the $T=0$~K case. The optical
conductivity remains the same as in the perfectly nested case for
$2\epsilon_0<\Delta(0,\epsilon_0)=\Delta_{00}$. 
For higher $\epsilon_0$ the optical conductivity is zero for $\omega<G$, $G=\Delta(\sqrt{8-a^2}-a)/2$
similarly to the effect of magnetic field where the $\omega<2\mu_BH$ part of
the conductivity is chopped\cite{tesla}, in other words a clean 
optical gap develops. This can readily be observed in Fig. \ref{fig:spec}: when the upper band crosses the 
Fermi energy, the chemical potential moves below the zeros of the gap, suppressing the low energy 
excitations, since only ${\bf q=0}$ transitions are allowed for.
Parallel to this the peak at $2\Delta$ splits into
2 new peaks at $\Delta(\sqrt{8-a^2}+a)/2$ and at $\Delta(a/2+2/a)$.
For $\omega>2\epsilon_0(1+a^2/4)$ the optical conductivity remains unchanged
compared to Ref. \onlinecite{nagycikk}. The only change in the remaining region can
be 
expressed by redefining the $I$ functions\cite{nagycikk}:
\begin{gather}
I(\alpha,\beta,g)=\frac{\omega^2 g}{4\Delta^2}\left(F\left(g\sqrt{\beta},x\right)-F\left(g\sqrt\alpha 
,x\right)
-E\left(g\sqrt\beta,x\right)+E\left(g\sqrt\alpha ,x\right)\right)\\
I_{sin}(\alpha,\beta,g)=\frac{\omega}{12\Delta}\left(\sqrt{\beta(1-\beta)\left(\left(\frac{\omega}{\Delta}\right)^2-4\beta^2\right)}-
\sqrt{\alpha(1-\alpha)\left(\left(\frac{\omega}{\Delta}\right)^2-4\alpha^2\right)}+\right.\nonumber \\
\left.
+\left(\frac{\omega}{\Delta g}+\frac 12\left(\frac{\omega 
g}{\Delta}\right)^3\right)\left(F\left(g\sqrt{\beta},x\right)-F\left(g\sqrt\alpha,x\right)\right)
-
\left(\frac{2\omega g}{\Delta}+\frac g2 \left(\frac 
\omega\Delta\right)^3\right)
\left(E\left(g\sqrt\beta,x\right)+E\left(g\sqrt\alpha ,x\right)\right)\right)
\end{gather}
and $I_{cos}(\alpha,\beta,g)=I(\alpha,\beta,g)-I_{sin}(\alpha,\beta,g)$, where $F(z,k)$
and $E(z,k)$ are the incomplete elliptic integrals of the first and second kind, $x=2\Delta/\omega g^2$
and the arguments of the $I$ functions are obtained as
\begin{eqnarray}
\alpha=\textmd{max}\left(0,{\frac 12-\frac{a^2}{8}-\frac{a\omega}{4\Delta}}\right)\\
\beta=\textmd{min}\left(1,{\frac 
12-\frac{a^2}{8}+\frac{a\omega}{4\Delta}},\left(\frac{\omega}{2\Delta}\right)^2\right)\\
g=\textmd{max}\left(1,\frac{2\Delta}{\omega}\right)
\end{eqnarray}
for $\omega>G$ and for $a<2$. For $a>2$, $\alpha=0$, $\beta=1$ and the
$I$ functions reduce to those in Ref. \onlinecite{nagycikk}. Here min and max gives the largest and the smallest 
value of its arguments, respectively. With these notations the optical conductivity reads as
\begin{eqnarray}
\textmd{Re}\sigma_{yy}^{sin,cos}(\omega)=e^2 N_0v_y^2\frac{8\Delta(0,\epsilon_0)^2}
{\omega^3}I_{sin,cos}\left(\alpha,\beta,g\right),\\
\textmd{Re}\sigma_{zz}(\omega)=e^2 N_0v_z^2\frac{4\Delta(0,\epsilon_0)^2}
{\omega^3}I\left(\alpha,\beta,g\right).
\end{eqnarray}

\begin{figure}[h!]
\psfrag{x}[t][b][1][0]{$\omega/\Delta_{00}$}
\psfrag{y}[b][t][1][0]{Re$\sigma_{zz}(\omega)\Delta_{00}/e^2N_0 v_z^2$}
{\includegraphics[width=11cm,height=7cm]{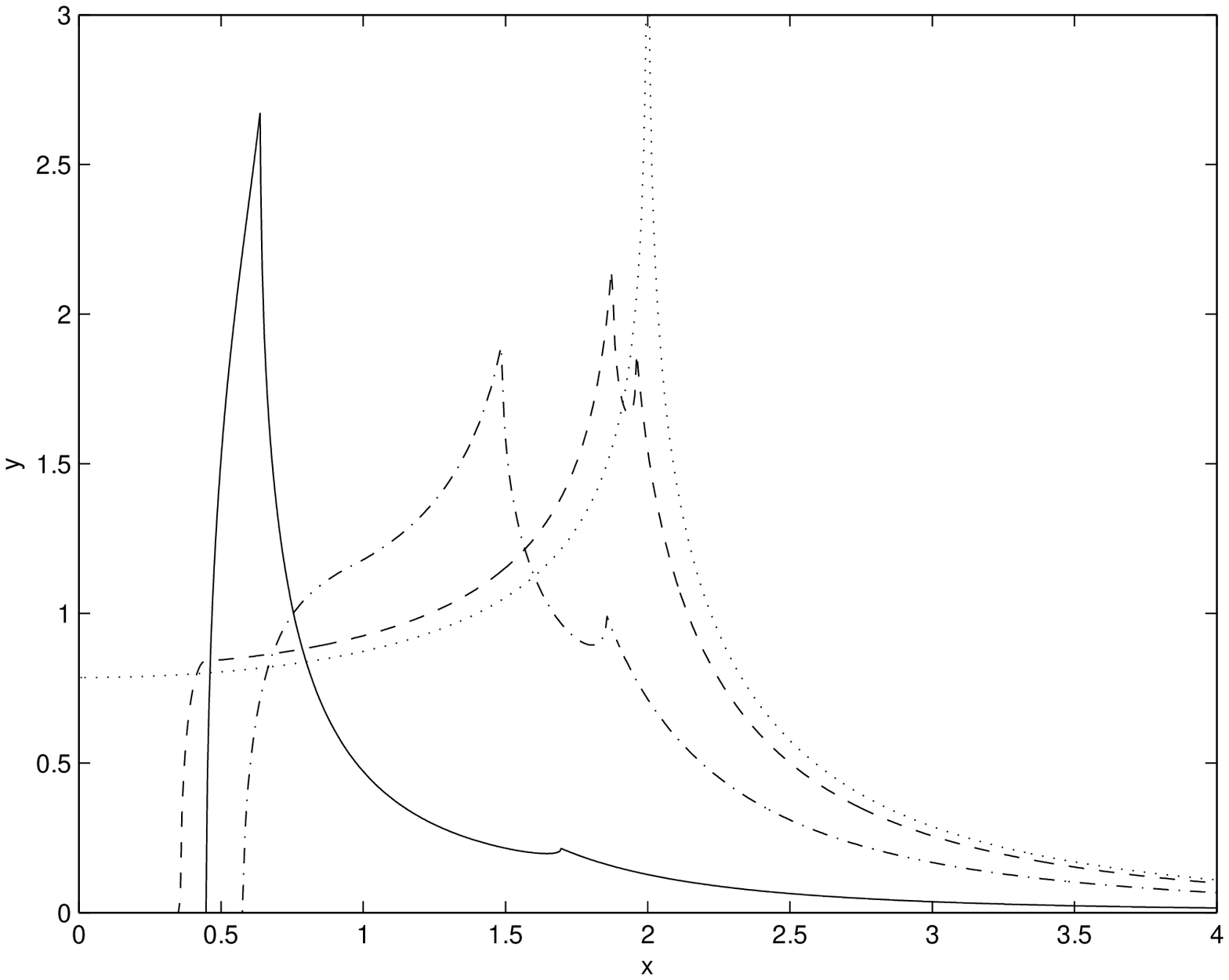}}

\vspace*{-6.7cm}\hspace*{63mm}
\psfrag{x}[t][b][1][0]{$\epsilon_0/\Delta_{00}$}
\psfrag{y}[b][t][1][0]{$G/\Delta_{00}$}
{\includegraphics[width=4cm,height=4cm]{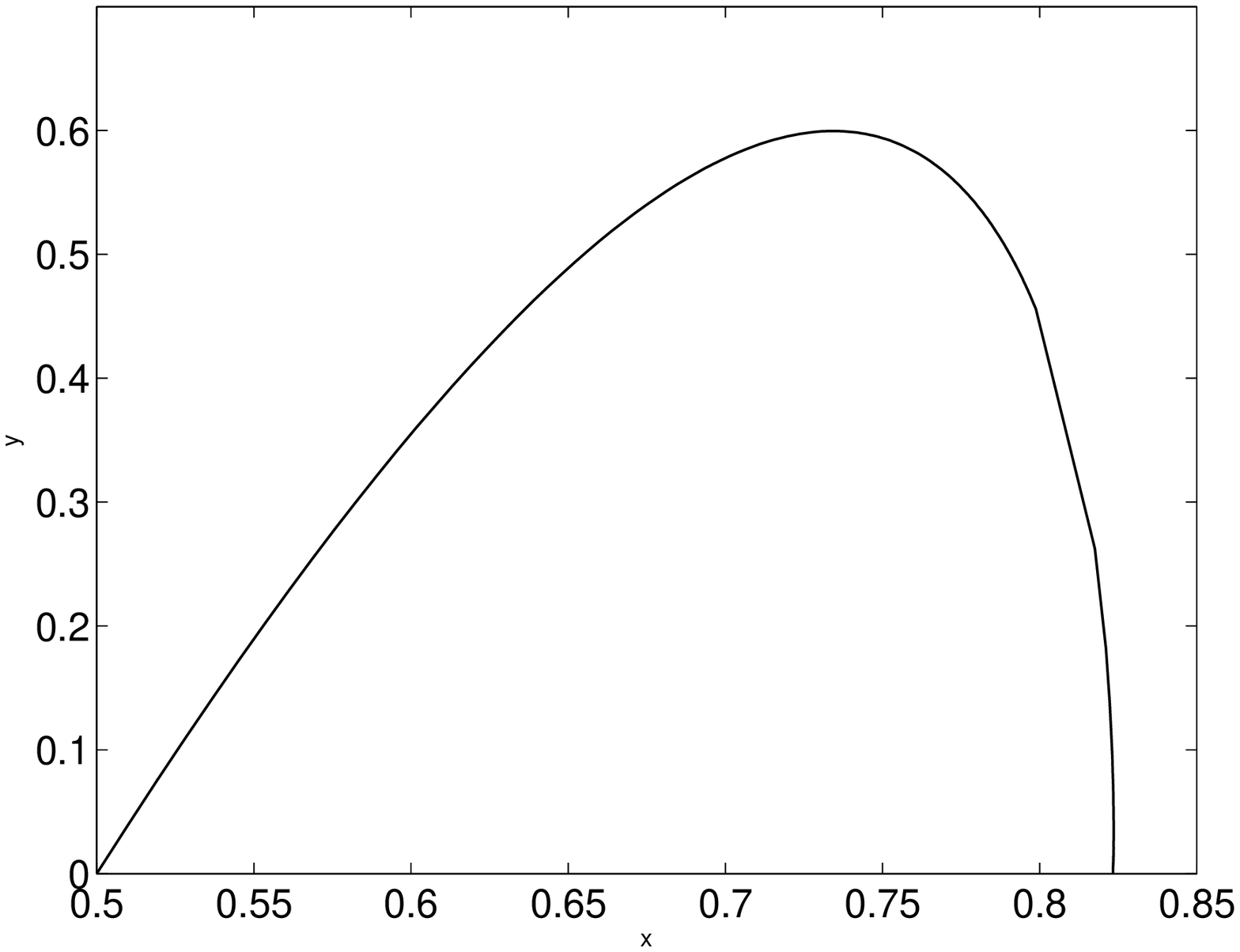}}

\vspace*{27mm}
\caption{The real part of the complex conductivity for the 2D model in the $z$ 
direction is shown 
for $\epsilon_0/\Delta_{00}=0-0.5$ (dotted line), $0.6$ (dashed line), 
$0.7$ (dashed-dotted line) and $0.8$ (solid line). Note that the same curves
belong to $\sigma_{xx}(\omega)$ by changing $v_z$ to $v_F$. The inset
shows the $\epsilon_0$ dependence of the optical gap.
\label{fig:vxx2}}
\end{figure}
\begin{figure}[h!]
\psfrag{x}[t][b][1][0]{$\omega/\Delta_{00}$}
\psfrag{y}[b][t][1][0]{Re$\sigma_{yy}^{sin}(\omega)2\Delta_{00}/e^2N_0 v_y^2$}
{\includegraphics[width=11cm,height=7cm]{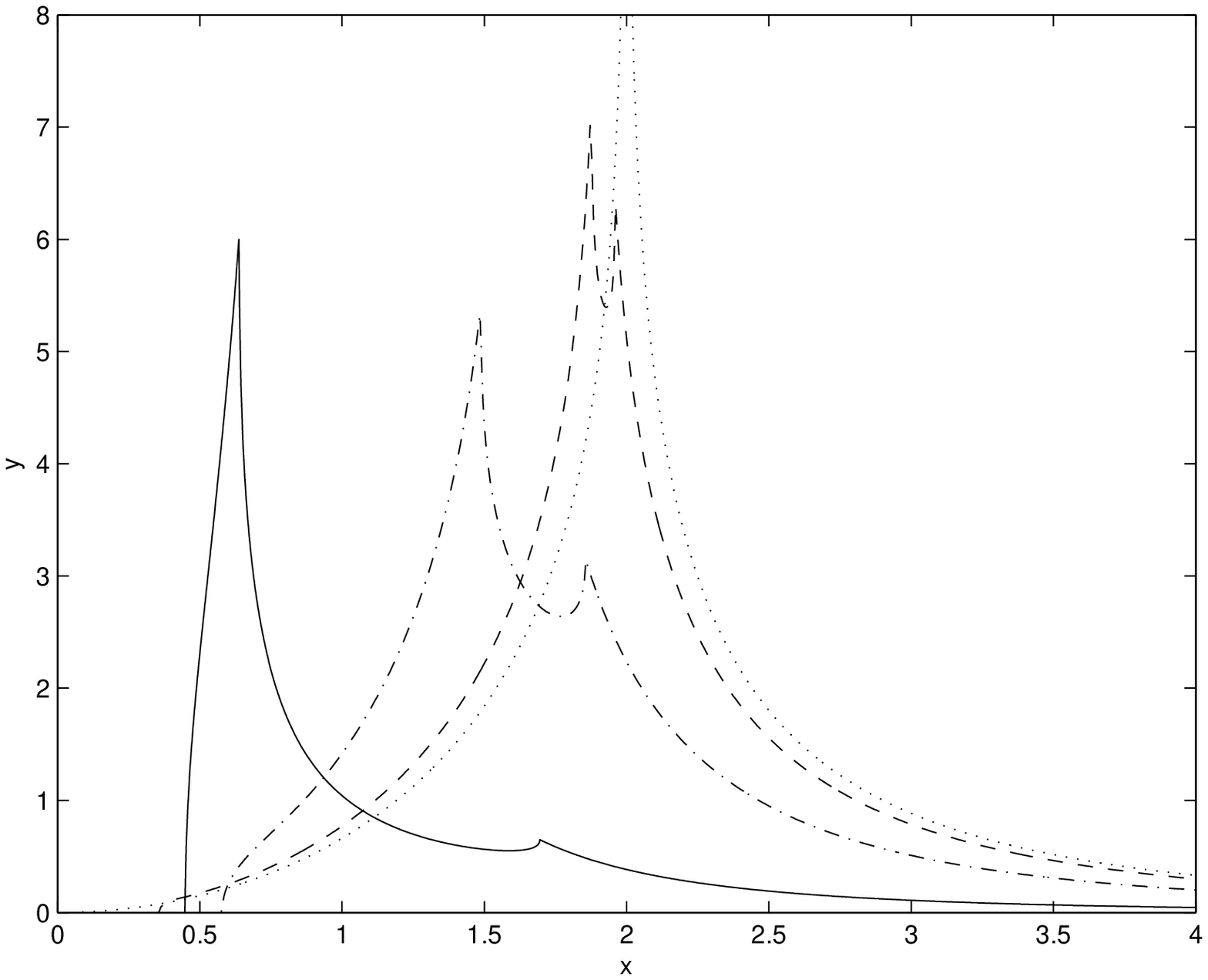}}

\vspace*{-6.7cm}\hspace*{63mm}
\psfrag{x}[t][b][1][0]{$\epsilon_0/\Delta_{00}$}
\psfrag{y}[b][t][1][0]{peak position}
{\includegraphics[width=4cm,height=4cm]{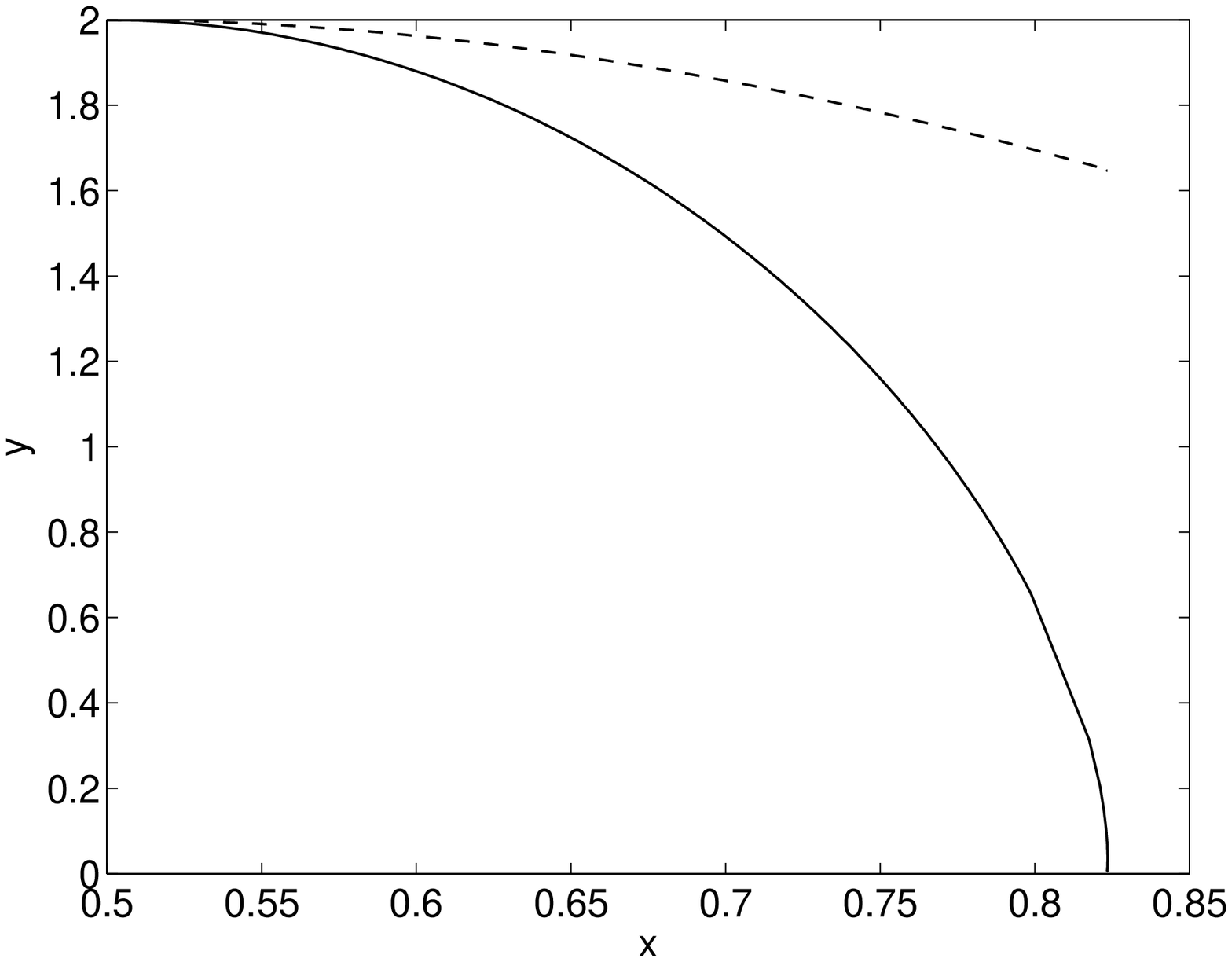}}

\vspace*{27mm}
\caption{The real part of the
complex conductivity for the 2D model for a
sinusoidal gap in the $y$ direction is shown 
for $\epsilon_0/\Delta_{00}=0-0.5$ (dotted line), $0.6$ (dashed line), 
$0.7$ (dashed-dotted line) and $0.8$ (solid line). The inset
shows the $\epsilon_0$ dependence of the peaks.\label{fig:vyys2}}
\end{figure}
\begin{figure}[h!]
\psfrag{x}[t][b][1][0]{$\omega/\Delta_{00}$}
\psfrag{y}[b][t][1][0]{Re$\sigma_{yy}^{cos}(\omega)2\Delta_{00}/e^2N_0 v_y^2$}
\centering{\includegraphics[width=11cm,height=7cm]{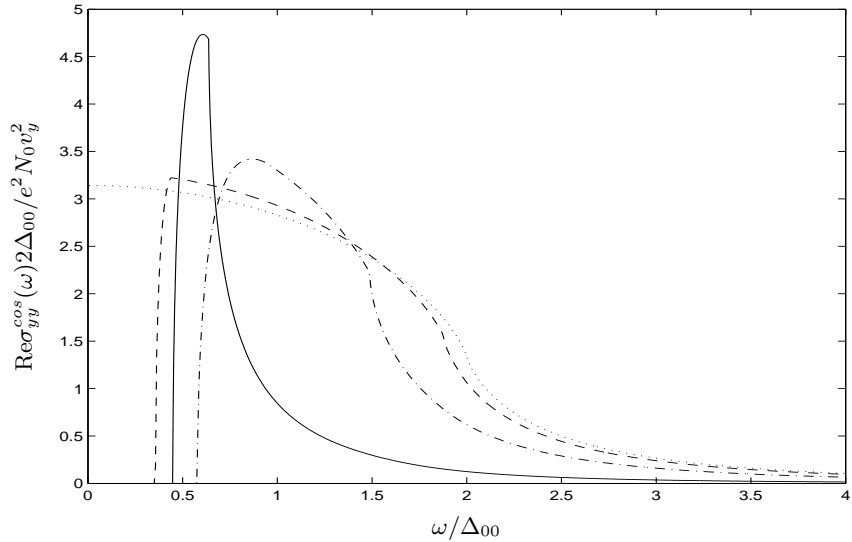}}
\caption{The real part of the
complex conductivity for the 2D model for a
cosinusoidal gap in the $y$ direction is shown 
for $\epsilon_0/\Delta_{00}=0$ (dotted line), $0.6$ (dashed line), 
$0.7$ (dashed-dotted line) and $0.8$ (solid line).\label{fig:vyyc2}}
\end{figure}

The optical conductivity in the three qualitatively different 
cases is shown in Figs. \ref{fig:vxx2},
\ref{fig:vyys2}, \ref{fig:vyyc2}. In the $x$ direction the quasiparticle
part of the optical conductivity 
is the same as $\sigma_{zz}(\omega)$ if we replace $v_z$ with $v_F$,
although in the $x$ direction
it does not give the total conductivity since collective contributions change significantly the
quasiparticle part as it was shown in Ref. \onlinecite{rpa}. At first sight the sum rule seems to be 
violated since a lot of optical
weight is missing at small
frequencies below the optical gap. But the $\delta(\omega)$ part of
the conductivity does not freeze out at $T\rightarrow 0$ in the presence of imperfect nesting,
and its coefficient provides the missing area.
As is well known, in the presence of impurity scattering, $\delta(\omega)$
changes to a Drude like peak centered at $\omega=0$.
At finite temperature, the optical gap vanishes, but excitations below $G$ are only possible with a 
probability of $\sim\exp(-(\epsilon_0-\Delta^2/(4\epsilon_0))/T)$.

\section{Conclusion}

We have studied theoretically the effect of imperfect nesting in unconventional density
waves. Two qualitatively different cases are
possible: the gap and imperfect nesting depend on the same (called 2D model) or different
wave vector component (3D case)\cite{imperfect}. Here we concentrated on the former. 
We explored the phase diagram which is identical to the one
in conventional density wave. The zero temperature gap  coefficient is not constant contrary to the 
conventional case. The chemical potential
changes compared to the normal state value. The density of states turned
out to be asymmetric with respect to the Fermi energy due
to the particle-hole symmetry breaking, but the logarithmically divergent
peaks of the $\epsilon_0=0$ case remain present, but at different
energies. For larger values of imperfect nesting
($2\epsilon_0>\Delta(T,\epsilon_0)$), the zero at the Fermi energy
disappears, and the low energy density of states regains its normal state form.
Usually $\epsilon_0$ is thought to vary with
pressure providing the opportunity to check these result in a wide range of
parameters. 
 The optical gap of the model in the perpendicular optical conductivity can
be observed  
experimentally at 
low temperatures. Moreover the splitting and lowering of the resonant peak at
$\omega=2\Delta$ (when the wavevector dependence of the gap and the velocity
coincide) or its absence (for the other kind of gap)\cite{imperfect} could provide robust signatures of the 
microscopic nature of
the low temperature phase.

\begin{acknowledgments}
This work was supported by the Magyary Zolt\'an postdoctoral
program of Foundation for Hungarian Higher Education and Research (AMFK) and by
the Hungarian
Scientific Research Fund under grant numbers OTKA TS040878, T046269 and NDF45172.
\end{acknowledgments}

\bibliography{eth}
\end{document}